\documentclass[%
 reprint,
 amsmath,amssymb,
 aps,
]{revtex4-1}

\usepackage{graphicx, amsmath,amssymb}% Include figure files
\usepackage{dcolumn}% Align table columns on decimal point
\usepackage{bm}% bold math
\begin{document}

%\preprint{APS/123-QED}

\title{Chaotic advection and mixing by a pair of rotlets in a circular domain}% Force line breaks with \\

\author{Phanindra Tallapragada}
  \email{ptallap@clemson.edu}%Lines break automatically or can be forced with \\
\author{Senbagaraman Sudarsanam}%
% \altaffiliation{Mechanical Engineering, Mahindra Ecole'}\\
 \email{ssudars@g.clemson.edu}
\affiliation{ Department of Mechanical Engineering, Clemson University
}%

\date{\today}% It is always \today, today,

\begin{abstract}
In this work we study chaotic mixing induced by point micro-rotors in a bounded two dimensional Stokes flow. The dynamics of the pair of rotors, modeled as rotlets, are non Hamiltonian in the bounded domain and produce chaotic advection of fluid tracers in subsets of the domain. A complete parametric investigation of the fluid mixing as a function of the initial locations of the rotlets is performed based on pseudo phase portraits. The mixing of fluid tracers as a function of relative positions of micro-rotors is studied using finite time entropy and locational entropy. The finite time locational entropy is used to identify regions of the fluid that produce good vs poor mixing and this is visualized by the stretching and folding of blobs of tracer particles.  Unlike the case of the classic blinking vortex dynamics, the velocity field of the flow modeled using rotlets inside a circular boundary is smooth in time and satisfies the no slip boundary condition. This makes the considered model a more realistic case for studies of mixing in microfluidic devices using magnetic actuated microspheres.
\begin{description}
\item[Usage]
Secondary publications and information retrieval purposes.
\item[PACS numbers]
May be entered using the \verb+\pacs{#1}+ command.
\item[Structure]
You may use the \texttt{description} environment to structure your abstract;
use the optional argument of the \verb+\item+ command to give the category of each item. 
\end{description}
\end{abstract}

\pacs{Valid PACS appear here}% PACS, the Physics and Astronomy
\maketitle

%\tableofcontents

\section{Introduction}\label{sec:intro}
Mixing and transport of fluids in the low Reynolds number regime is an important challenge that has gained considerable attention over the years  \cite{Ottino2004,Stremler2004,Ward2015}. As a consequence of weak fluid inertia, viscous diffusion and chaotic advection are the dominant modes of mixing in low Reynolds number flows. Viscous diffusion unfolds over microscopic length scales and by itself is insufficient to achieve fast mixing in most microfluidic applications which require rapid mixing. This underscores the importance of advective effects for stretching and folding of fluid material lines that in turn enhance viscous diffusion by increasing the area of contact between fluids being mixed. Various means of achieving rapid advective microfluidic mixing have been proposed over the years \cite{Lee2011} which can be classified broadly into active and passive mixing methods. Active mixing mechanisms induce large scale motion of the fluid by directly influencing the fluid by means of mixing elements driven by force fields such as electric/magnetic or acoustic fields that compensate for the lack of fluid inertia \cite{Bertin2017,Xia2018,Wu2017} and can achieve high mixing quality especially in the vicinity of the mixing elements. Passive methods of mixing involve altering the boundary conditions of the flow so as to maximize the area of contact between fluids being mixed. This is usually achieved by incorporating geometric features in the path of the fluid flow \cite{Stroock2002} or by altering the path itself such that fluid stretches and folds into finer scale structures \cite{Sudarsan2006}. Passive methods are high throughput methods and achieve a more uniform mixing than active methods. The quality of passive mixing however, is directly dependent on the duration of time the fluid spends within the mixing geometry necessitating the need for longer microfluidic channels. %In this work, we study a particularly simple model for active micro-mixers which are not localized in space and thus achieve efficient and homogeneous mixing at fast time scales.

Naturally occurring micro-swimmers have been known to induce mixing of the surrounding fluid during locomotion \cite{Kurtuldu2011,Sokolov2009}. They accomplish this by continuously stirring the fluid around them as they swim, thus making up for the weak inertia of the fluid. Inspired by these studies, artificial micro-swimmers have gained attention as a means to mix and predictably manipulate fluids at small scales \cite{Gilbert2011,Jalali2015}. While many different micro-swimmer morphologies have been proposed in these works and others, one of the simplest synthetic micro-swimmers that can be studied for this purpose is a self propelled spinning particle. Grzybowski demonstrated the self-assembly of spinning rotors in a plane \cite{Grzybowski2000,Grzybowski2002} and later Campbell and Grzybowski used these ideas to experimentally demonstrate the possibility of mixing using magnetically driven rotors \cite{Campbell2004}. More recently, Ballard et.al  \cite{Ballard2016} demonstrated microfluidic mixing due to magnetic micro-beads using experiments and lattice Boltzmann simulations. These and other experimental investigations \cite{Lee2009} of magnetic or otherwise driven-particles have shown their effectiveness for micro-mixing applications.

The dynamics of such spinning particles or 'micro-rotors' were recently studied numerically \cite{Lushi2015,Yeo2015} where the flow generated by the micro-rotors were modeled using a rotlet velocity field. The rotlet also known as a couplet is a solution to the stokes equation with a point torque inhomogeneity  \cite{Chwang1974,Batchelor1971}. Meleshko and Aref  \cite{Meleshko1996} showed that \textit{blinking} rotlets under circular confinement could be an effective model to study mixing in viscous flows where they present the blinking rotlet as a viscous analogue of the paradigmatic blinking point vortex in inviscid flows pioneered by Aref \cite{Aref1984}.  In these studies the location of the blinking singularities are fixed in the domain and the singularities themselves do not interact with each other. 

In the present paper we investigate the mixing of the fluid induced in a circular domain due to the dynamic evolution of the positions of a pair of interacting rotlets. A pair of interacting rotlets can exhibit rich dynamics in a circular domain due to the effects of the boundary. In the absence of a boundary, the dynamics of interacting rotlets has a Hamiltonian structure with linear and angular momentum being integrals of motion. In a circular domain, the linear and angular momentum are no longer conserved and their motion is non integrable. This is in contrast to the problem of two point vortices in a circular domain, which is known to be integrable and produces quasiperiodic motion of the vortices, \cite{Kimura1988}. Nevertheless the dynamics of the two rotlets are nearly Hamiltonian. Furthermore the dynamics of a passive tracer induced by the pair of rotlets is the restricted three rotlet problem and exhibits chaotic motion. We investigate the dynamics of passive tracers advected by the flows generated by micro rotors using finite time locational entropy. We  show that the nearly Hamiltonian structure of the dynamics of the pair of rotlets can be used to classify rotlet motions that produce better good mixing of the fluid in the domain. The present study of mixing is semi active; a physical realization of the rotlets requires continuous actuation, say via a time periodic magnetic field that can exert a torque on magnetic spheres and keep the spheres spinning.  However the motion of the spinning spheres through the domain is purely due to the hydrodynamic interaction between the spheres and the boundary and mixing of the fluid that is induced does not have to directly controlled. Recent work has shown that a pair of magnetic spheres can be steered in a circular domain by a spatially uniform magnetic field as a single control input, \cite{bfst_dscc2018, bft_ijira2018} making the microrotors steerable mobile micromixers.   %We show that mixing as quantified by a finite time entropy increases when the motion of the rotlets departs significantly from being Hamiltonian. %The motion of the rotlets is such that fluid from the interior rotlet is stretched outwards and anchored to the boundary when the two rotlets pass each. The resulting stretching and folding produced by these slowly moving anchoring points, which are instantaneous stagnations points in the flow produce mixing of the fluid.  Mixing efficiency is quantified by  the decay of variance of an initial concentration field.

\section{Dynamics of Rotlets }
The flow due to a spinning micro particle (shaped as flat disc or a sphere) can be modeled as a singular torque in a quiescent, viscous fluid. In the vanishing Reynolds number limit, the governing equation of motion for fluid is the Stokes equation. In this regime a micro particle spinning with a constant angular velocity $\omega$ in an otherwise quiescent fluid requires a constant torque to balance the viscous hydrodynamic torque. The disturbance velocity field in the plane of the particle generated by such a spinning micro particle can be modeled by that of a Stokes singularity called the rotlet \cite{Pozrkidis1992}. This disturbance velocity is given by
%\begin{eqnarray}
%\mu\nabla^2u=\nabla\times\gamma\delta(\mathbf{x-x_0}) \\
%\nabla\cdot u=0\nonumber .
%\end{eqnarray}
\begin{equation}
u(\mathbf{x})=-\gamma \hat{k}\times\frac{\mathbf{x-x_0}}{||\mathbf{x-x_0}||^2} .
\end{equation}
for a rotlet located $\mathbf{x_0} = (x_0,y_0)$, where $\gamma$ is the strength of the rotlet with the associated stream function being $\psi(x,y) = \gamma\ln |\mathbf{x}-\mathbf{x_0}|^2$. The velocity field generated by the two dimensional rotlet in an unbounded domain is topologically the same as that generated by an ideal point vortex. Analogous to the classical point vortex which can be thought of as the intersection of a 3D vortex filament and a plane perpendicular to it \cite{Aref2007}, a point rotlet can be considered to be the intersection of an infinitesimal spinning particle and a horizontal plane passing through its center. In the absence of any boundaries or other singularities in its vicinity, the rotlet is fixed in position and the flow is associated with circular streamlines around the location of the singularity.

In an assembly of $N$ rotlets in a plane with locations $(x_i, y_i)$ for $i \in [1,N]$, each rotlet is advected by a velocity that is a linear superposition of the velocities induced by all the other rotlets.  The motion of the $i$-th  rotlet can then be expressed by,

\begin{eqnarray}\label{eq:rotlet_sum}
\frac{dx_i}{dt}=\sum_{j=1,j\neq i}^{j=N}\gamma_j\frac{(y_i-y_j)}{R_{ij}^2} \nonumber \\
\frac{dy_i}{dt}=\sum_{j=1,j\neq i}^{j=N}-\gamma_j\frac{(x_i-x_j)}{R_{ij}^2}
\end{eqnarray}
where, the square of distance between the $i$-th and $j$-th rotlet is given by $R_{ij}^2=(x_i-x_j)^2+(y_i-y_j)^2$.

The interaction of $N$ rotlets in an unbounded domain has a Hamiltonian structure and conserves the linear and angular impulse \cite{Lushi2015}. These integrals of motion preclude chaotic motion in the dynamics of three of fewer rotlets. This also precludes the chaotic advection of a fluid by the motion of two rotlets. This is because the interaction of a two rotlets and a tracer can be treated as a restricted three-rotlet problem with the third rotlet having zero strength.   However mixing of fluid is usually performed in a bounded domain. We show that when the fluid domain is bounded, as it is in practical applications, the motion of two rotlets can produce chaotic advection of the fluid. Specifically we consider the mixing of fluid due to the dynamics of two rotlets confined to the interior of a unit circle.

%\subsection{Dynamics of rotlets in a circular domain}\label{sec:eqs}
The dynamics of rotlets in a circular domain of radius $a$, differ significantly compared to those in an unbounded domain. The velocity of the fluid on the circular boundary has to be zero. These boundary conditions can be satisfied by the method of images. The stream function governing the motion of the fluid, due to the presence of a rotlet of strength $\gamma_1$ located at $(x_1,y_1)$, such that $\sqrt{x_1^2+y_1^2} = r_1 \leq a$ derived by Ranger \cite{Ranger1980} and Aref and Meleshko \cite{Meleshko1996} is,
\begin{equation}\label{eq:psi}
    \psi_1(x,y) = \gamma_1\left(\ln A_1(x,y) - \ln B_1(x,y) + \frac{C_1(x,y)}{B_1(x,y)}\right)
\end{equation}
    where
\begin{align}
    A_1(x,y) &= (x-x_1)^2+(y-y_1)^2 \nonumber\\
    B_1(x,y) &= \left[\left(x-x_1\frac{a^2}{r_1^2} \right)^2 + \left(y-y_1\frac{a^2}{r_1^2} \right)^2\right]\frac{r_1^2}{a^2} \nonumber\\
    C_1(x,y) &= \left(1 - \frac{x^2+y^2}{a^2}\right) \left(a^2 - (x^2+y^2)\frac{r_1^2}{a^2} \right). \nonumber
\end{align}
The first term in \eqref{eq:psi} is associated with the velocity of the fluid due to a rotlet in an unbounded domain. The second part of the stream function $-\gamma_1 \ln{B_1(x,y)}$ ensures that the velocity of the fluid on the boundary has no normal component. This term is in fact the stream function due to an image vortex placed at a location outside the circle according to the Milne-Thomson circle theorem. The third term $\frac{C_1(x,y)}{B_1(x,y)}$ ensures that the zero slip condition on the boundary is satisfied. This is unique to the viscous flow setting and leads to significant differences in the motion of rotlets inside the circle as compared to that of  point vortices.

When $N$ rotlets are present in a circular region, each rotlet experiences a velocity due to the other $N-1$ rotlets and their image systems and its own image system, 
\begin{align} \label{eq:rotlet_circle}
    \dot{x}_i &= \sum_{j=1, j\neq i}^N\gamma_j\frac{\partial}{\partial y_i} \ln A_j(x_i,y_i) + \sum_{j=1}^N-\gamma_j\frac{\partial}{\partial y_i} \ln B_j(x_i,y_i) \nonumber \\ &+ \sum_{j=1}^N \gamma_j\frac{\partial}{\partial y_i}  \left(\frac{C_j(x_i, y_i)}{B_j(x_i,y_i)}\right) \nonumber \\
    \dot{y}_i &= \sum_{j=1, j\neq i}^N-\gamma_j\frac{\partial}{\partial x_i} \ln A_j(x_i,y_i) + \sum_{j=1}^N\gamma_j\frac{\partial}{\partial x_i} \ln B_j(x_i,y_i) \nonumber \\ &-\sum_{j=1}^N \gamma_j\frac{\partial}{\partial x_i}  \left(\frac{C_j(x_i, y_i)}{B_j(x_i,y_i)}\right).
\end{align}
The velocity of a rotlet is a linear combination of two components. One component of this velocity (the terms containing  $\ln A_j$ and $\ln B_j$), denoted by $(\dot{x}_{iv}, \dot{y}_{iv})$  is similar to the velocity experienced by point vortices and the second component is a velocity induced by the zero slip velocity condition (the terms containing  $\frac{C_j}{B_j}$, denoted here after as $(\dot{x}_{is}, \dot{y}_{is})$. Equation \eqref{eq:rotlet_circle} can be rewritten as 
\begin{align} \label{eq:rotlet_eqs2}
    \dot{x}_i &= \dot{x}_{iv} + \dot{x}_{is} \nonumber \\
    \dot{y}_i &= \dot{y}_{iv} + \dot{y}_{is}.
\end{align}
Two special cases of integrable Hamiltonian dynamics arise from \eqref{eq:rotlet_eqs2}. The first is the simple case of $N=1$. In this case the Hamiltonian is $H_1 = -\gamma (\ln{B_1(x_1,y_1)}-\frac{C_1(x_1,y_1)}{B_1(x_1,y_1)})$ with $\dot{x}_1 = \frac{1}{\gamma_1}\frac{\partial H_1}{\partial y_1}$ and $\dot{y}_1 = -\frac{1}{\gamma_1}\frac{\partial H_1}{\partial y_1}$ the system is completely integrable. It can be verified that when a single rotlet is present in a circular domain, its distance from the center of the circle remains invariant. The second case of Hamiltonian dynamics arise when considering the motion of two point vortices instead of rotlets, i.e. $C(x,y) =0$, $\dot{x}_{is}=0$ and $\dot{y}_{is}=0$ (see \cite{Kimura1988}). A direct calculation shows that the angular impulse $I = \gamma_1 r_1^2 + \gamma_2 r_2^2$ is invariant, where $r_i^2 = x_i^2+y_i^2$. Therefore the motion of two point vortices in a circular domain is integrable \cite{Kimura1988}. %Following \cite{Kimura1988} we will express the Hamiltonian, $H$, in terms of the function $G(\theta, \eta) = exp(-4\pi H)$ where $\theta = \theta_1 - \theta_2$ and $\theta_1 = \tan^{-1}\left( \frac{y_1}{x_1}\right)$ and $\theta_2 = \tan^{-1}\left( \frac{y_2}{x_2}\right)$. The function $G(\theta, \eta)$ given by
%\begin{equation}\label{eq:G}
 %   G = \frac{((1-I)^2-\eta^2)(1+I^2-\eta^2-2\sqrt{I^2-\eta^2}\cos{\theta})}{2I-2\sqrt{I^2-\eta^2}\cos{\theta}}.
%\end{equation}
%The variable $\eta$ is defined as
%\begin{equation}\label{eq:eta}
 %   r_1^2 = I - \eta \;\;\;\; \text{and} \;\;\;\; r_2^2 =I +\eta.
%\end{equation}
%Kimura \cite{Kimura1988} showed three critical values of $I$ for the function $G$ \eqref{eq:G}. These occur at about $I_1=0.21374$, $I_2=0.29038$ and $I_3=0.5$.

%\begin{figure}[!h]
%\begin{minipage}{0.23\textwidth}
%\begin{center}
%		\includegraphics[width=0.95\textwidth]{contours_I01.eps}\\
%(a)$I=0.1 <I_1$\\
%\end{center}
%\end{minipage}
%\begin{minipage}{0.23\textwidth}
%\begin{center}
%		\includegraphics[width=0.95\textwidth]{contours_I025.eps}\\
%(b)$I_1<I=0.25<I_2$\\
%\end{center}
%\end{minipage}

%\begin{minipage}{0.23\textwidth}
%\begin{center}
%	\includegraphics[width=0.95\textwidth]{contours_I032.eps}\\
%(c)$I_2<I=0.32<I_3$
%\end{center}
%\end{minipage}
%\begin{minipage}{0.23\textwidth}
%\begin{center}
%		\includegraphics[width=0.95\textwidth]{contours_I07.eps}\\
%(d)$I_3<I=0.7$
%\end{center}
%\end{minipage}

%\caption{ Level sets of the function $G(\theta, \eta)$ showing qualitative changes as $I$ is varied.}\label{fig:G_contours}
%\end{figure}

Apart from these special cases, in the general case of the motion of two rotlets, the zero slip velocity condition on the circular boundary, no longer conserves the angular impulse. A direct computation using \eqref{eq:rotlet_eqs2} verifies that 
\[\gamma_1(x_1\dot{x}_{1s} + y_1\dot{y}_{1s})+\gamma_2(x_2\dot{x}_{2s} + y_2\dot{y}_{2s}) \neq 0.\]
It can be shown that equations \eqref{eq:rotlet_eqs2} are not Hamiltonian for $N>1$. For if a Hamiltonian exists $H(x_i,y_i)$, then $\dot{x}_i = \frac{1}{\gamma_i} \frac{\partial H}{\partial y_i}$ and $\dot{y}_i = -\frac{1}{\gamma_i}\frac{\partial H}{\partial x_i}$. This would imply the canonical $\frac{\partial \dot{x}_i}{\partial x_i} + \frac{\partial \dot{y}_i}{\partial y_i} =0 $ which \eqref{eq:rotlet_eqs2} satisfy. However the existence of a Hamiltonian would also imply that $\frac{\partial \dot{x}_i}{\partial y_j} - \frac{\partial \dot{x}_j}{\partial y_i} =0$. Because of the no slip condition on the boundary, $\frac{\partial \dot{x}_{is}}{\partial y_j} - \frac{\partial \dot{x}_{js}}{\partial y_i} \neq 0$ while $\frac{\partial \dot{x}_{iv}}{\partial y_j} - \frac{\partial \dot{x}_{jv}}{\partial y_i} = 0$. Therefore the system \eqref{eq:rotlet_eqs2} is not Hamiltonian. 

\begin{figure}[!h]
\begin{minipage}{0.15\textwidth}
\begin{center}
	\includegraphics[width=0.98\textwidth]{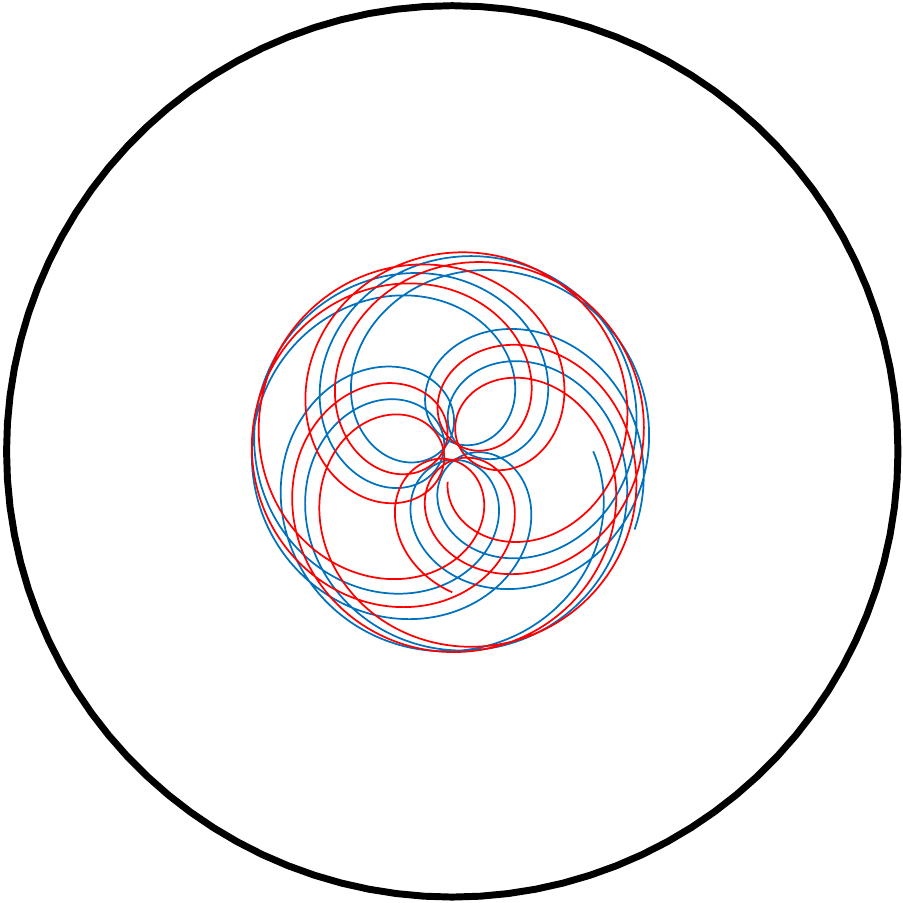}\\
(a)$I(0)=0.1$
\end{center}
\end{minipage}
\begin{minipage}{0.15\textwidth}
\begin{center}
		\includegraphics[width=0.98\textwidth]{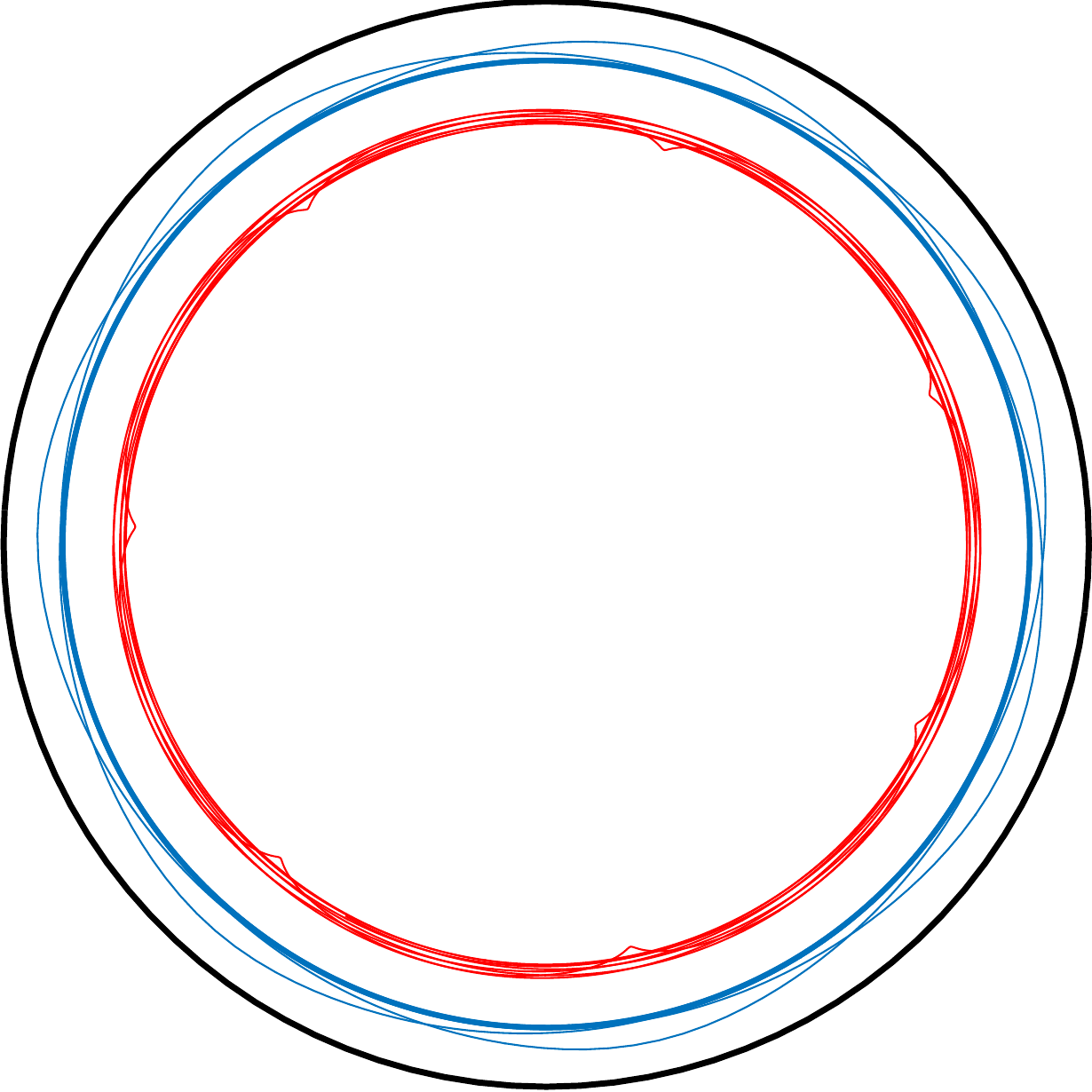}\\
(b)$I(0)=0.7$
\end{center}
\end{minipage}
\begin{minipage}{0.15\textwidth}
\begin{center}
		\includegraphics[width=0.98\textwidth]{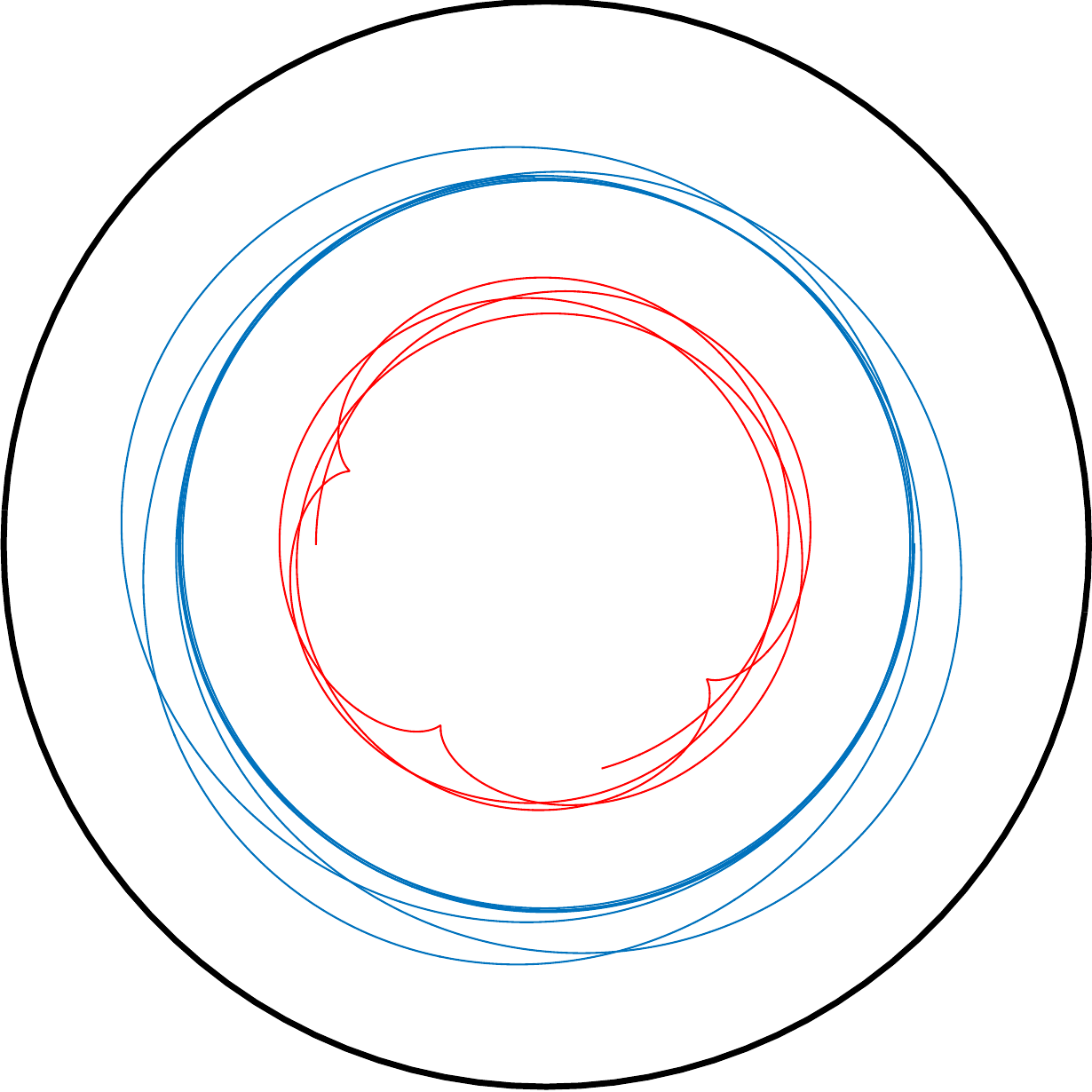}\\
(c)$I(0)=0.15$
\end{center}
\end{minipage}

\caption{Trajectories of the pair of rotlets for three sample initial conditions $(I(0), \eta(0)=0, \theta(0)=\pi/2)$, with $(x_1(0) = 0 , y_1(0)=\sqrt{I(0)})$ and $(x_2(0) = \sqrt{I(0)}, y_2(0)=0)$. (a) $(I(0)=0.1$,  (b) $I(0)=0.7$ and (c) $I(0)=0.0.32$.  }\label{fig:traj}
\end{figure}

While the system \eqref{eq:rotlet_eqs2} is not integrable when $N>1$, the dynamical system \eqref{eq:rotlet_eqs2} is nearly Hamiltonian in the sense that if $\dot{x}_{is} \ll \dot{x}_{iv}$ and $\dot{y}_{is} \ll \dot{y}_{iv}$, then $\dot{x}_i \approx \frac{1}{\gamma_i}\frac{\partial H}{\partial y_i}$ and $\dot{y}_i \approx -\frac{1}{\gamma_i}\frac{\partial H}{\partial x_i}$. This can occur in two scenarios. The first is when the effects of the boundary are `small', i.e. both the rotlets are far from the boundary, $r_i^2<<a^2$. In this case the dynamics of \eqref{eq:rotlet_circle} approximate the dynamics of two rotlets in an unbounded domain.  The second case occurs when the two rotlets are far from each other but close to the boundary. In this case the interaction between the rotlets is weak and the motion of each rotlet is almost decoupled from that of the other and the system is nearly Hamiltonian. This nearly Hamiltonian nature of the system can be used to parameterize the solutions of \eqref{eq:rotlet_eqs2} in terms of two variables $\eta$ defined as
\begin{equation}\label{eq:eta}
   r_1^2 = I - \eta \;\;\;\; \text{and} \;\;\;\; r_2^2 =I +\eta
\end{equation}
and  $\theta = \theta_1 - \theta_2$ and $\theta_1 = \tan^{-1}\left( \frac{y_1}{x_1}\right)$ and $\theta_2 = \tan^{-1}\left( \frac{y_2}{x_2}\right)$. Figure \ref{fig:traj} shows the trajectories of the two rotlets for three different initial conditions for $\gamma_1=\gamma_2=1$.    %This corresponds to the case $I>I_3$. %Figure \ref{} shows three sample trajectories of the two rotlets for three categories of $I$ shown in \ref{}. The function $G$ is also plotted in each of these cases showing small variations when $I<I_1$ and $I_3<I$ and thus the nearly Hamiltonian nature of the system \ref{eq:rotlet_eqs2} and large variations when $I_2<I<I_3$. These observations motivate classifying the trajectories of the rotlets  by the underlying Hamiltonian case to investigate the mixing of the fluid that their motion produces.

\begin{figure}[!h]
\begin{minipage}{0.23\textwidth}
\begin{center}
	\includegraphics[width=0.95\textwidth]{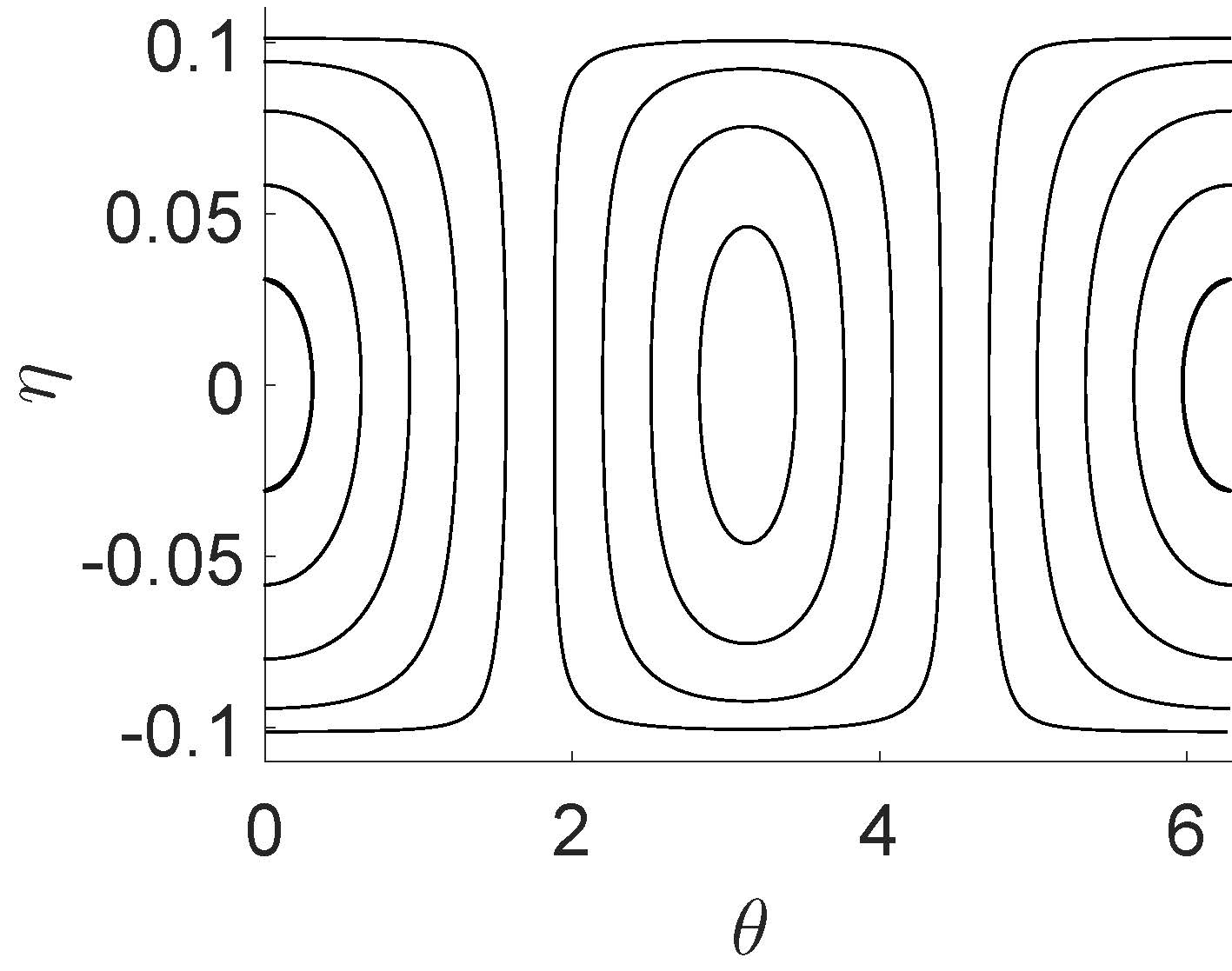}\\
(a)$I(0)=0.1$
\end{center}
\end{minipage}
\begin{minipage}{0.23\textwidth}
\begin{center}
		\includegraphics[width=0.95\textwidth]{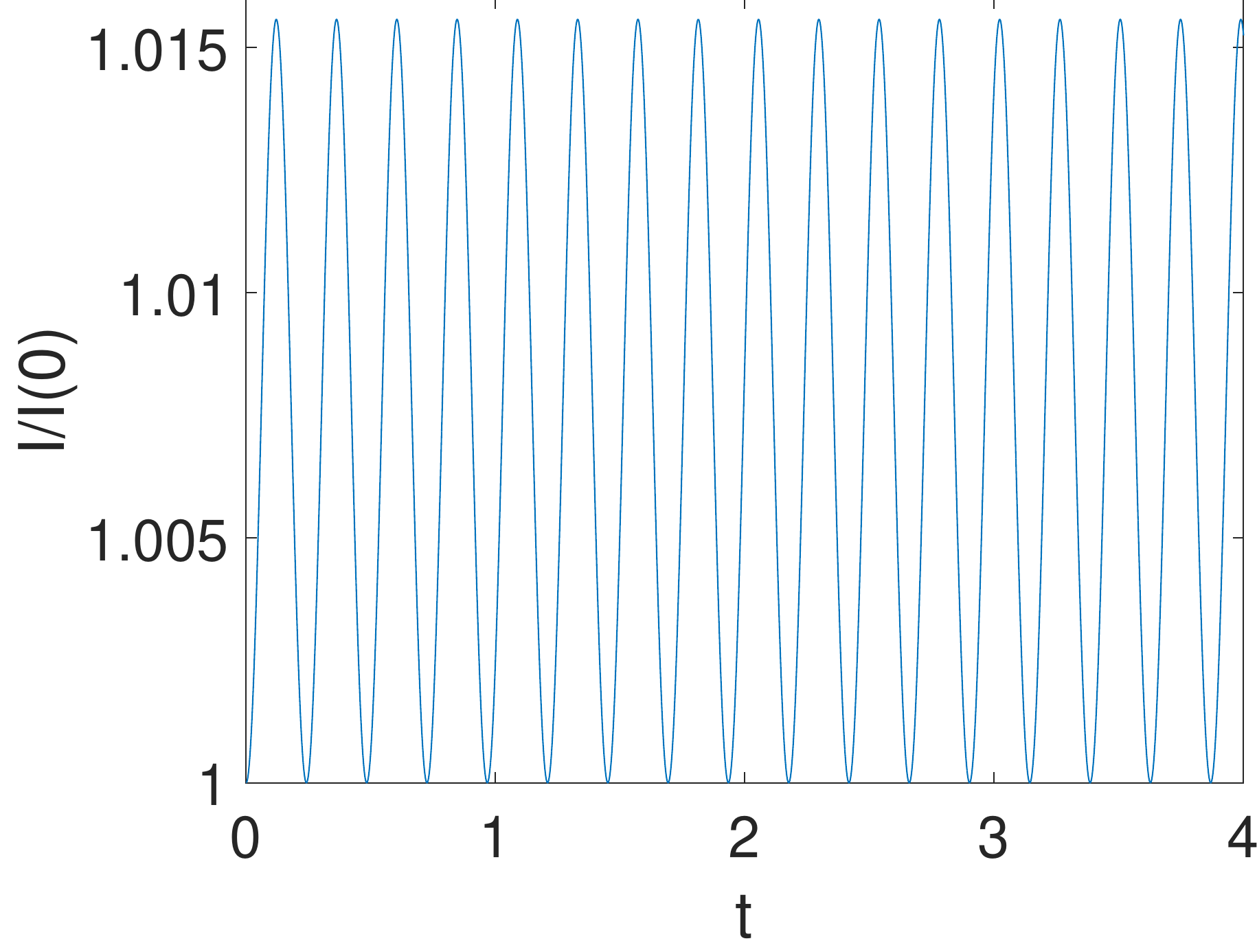}\\
(e)%$\frac{I(t)}{I(0)}$
\end{center}
\end{minipage}

\begin{minipage}{0.23\textwidth}
\begin{center}
	\includegraphics[width=0.95\textwidth]{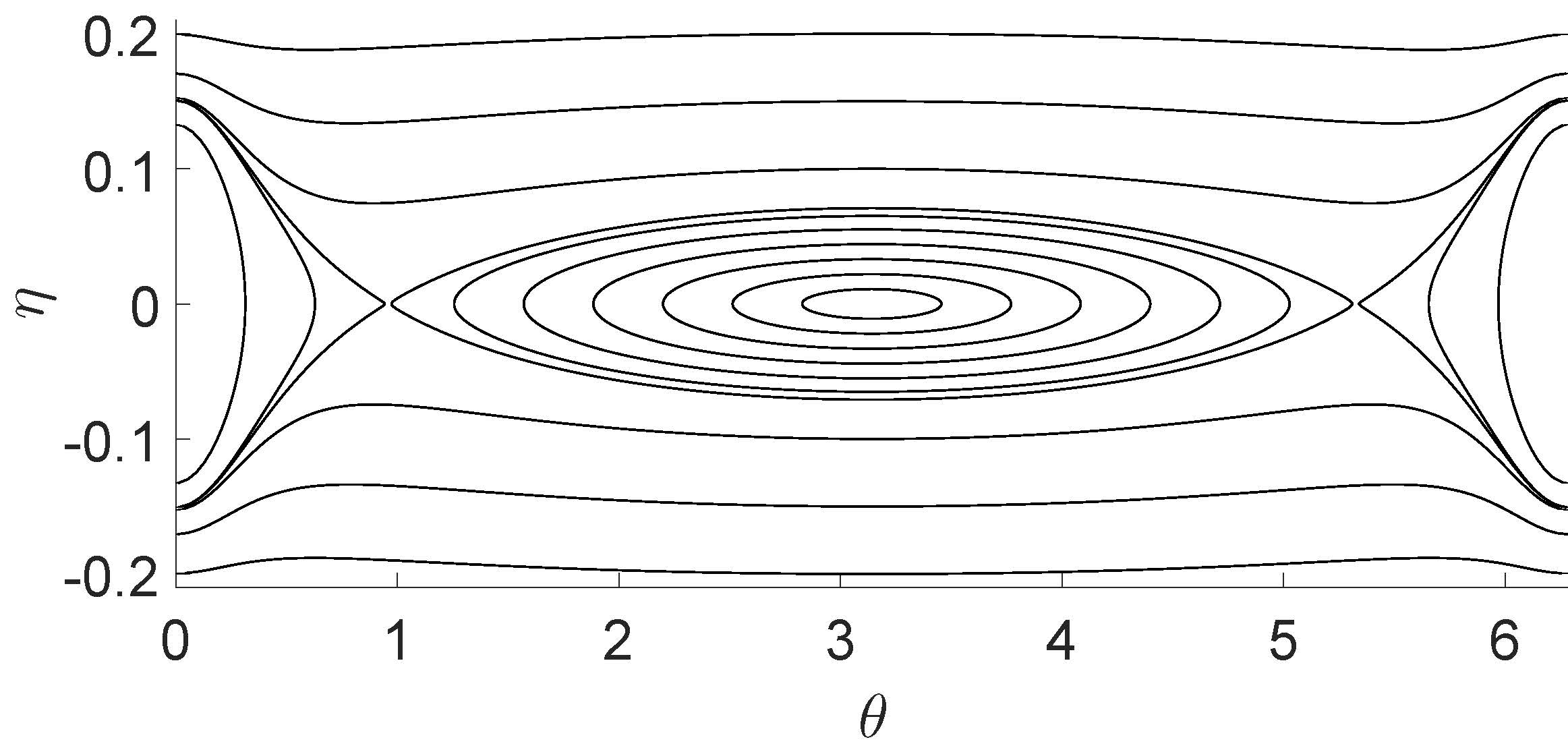}\\
(b)$I(0)=0.7$
\end{center}
\end{minipage}
\begin{minipage}{0.23\textwidth}
\begin{center}
		\includegraphics[width=0.95\textwidth]{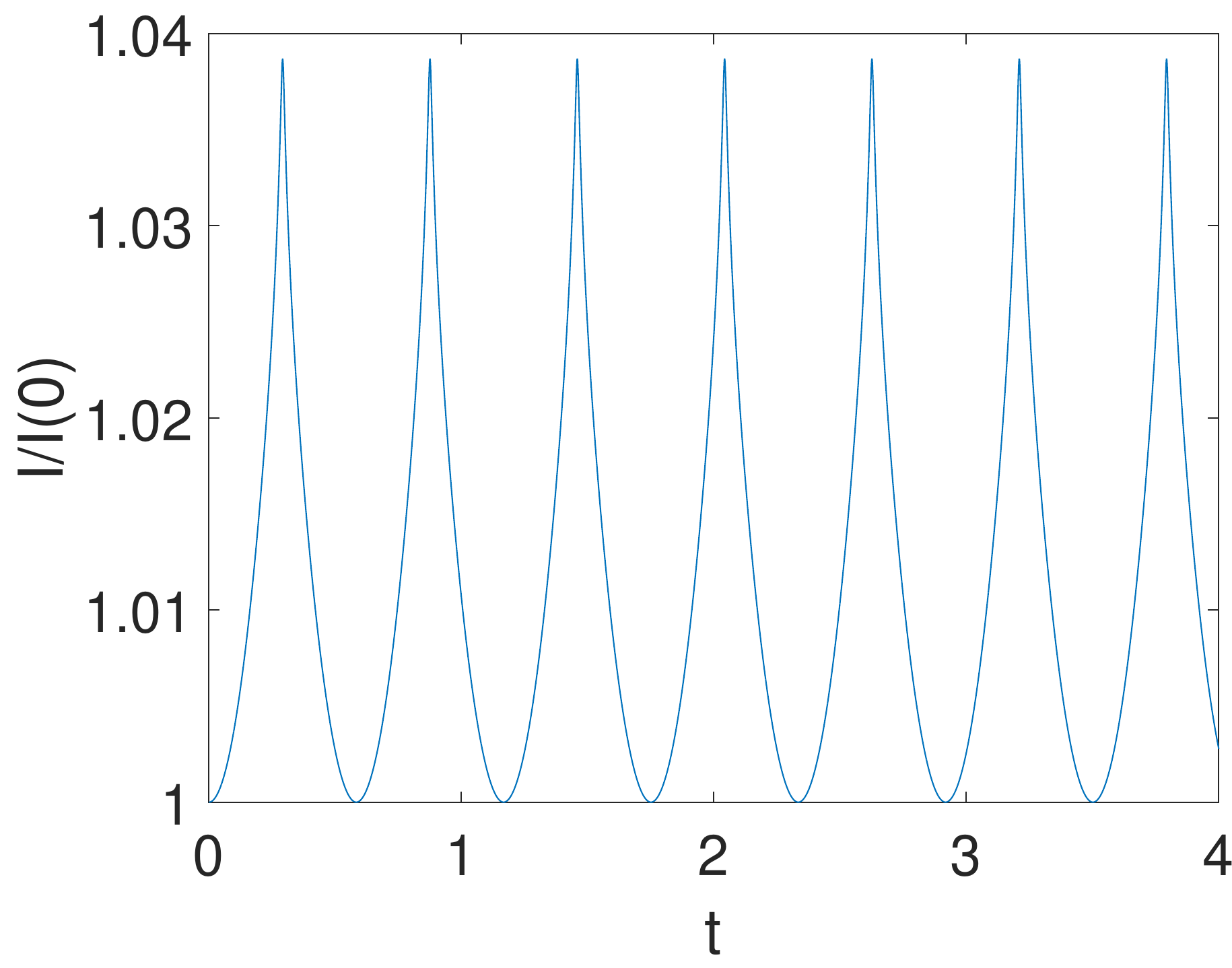}\\
(f)%$\frac{I(t)}{I(0)}$
\end{center}
\end{minipage}

\begin{minipage}{0.23\textwidth}
\begin{center}
	\includegraphics[width=0.95\textwidth]{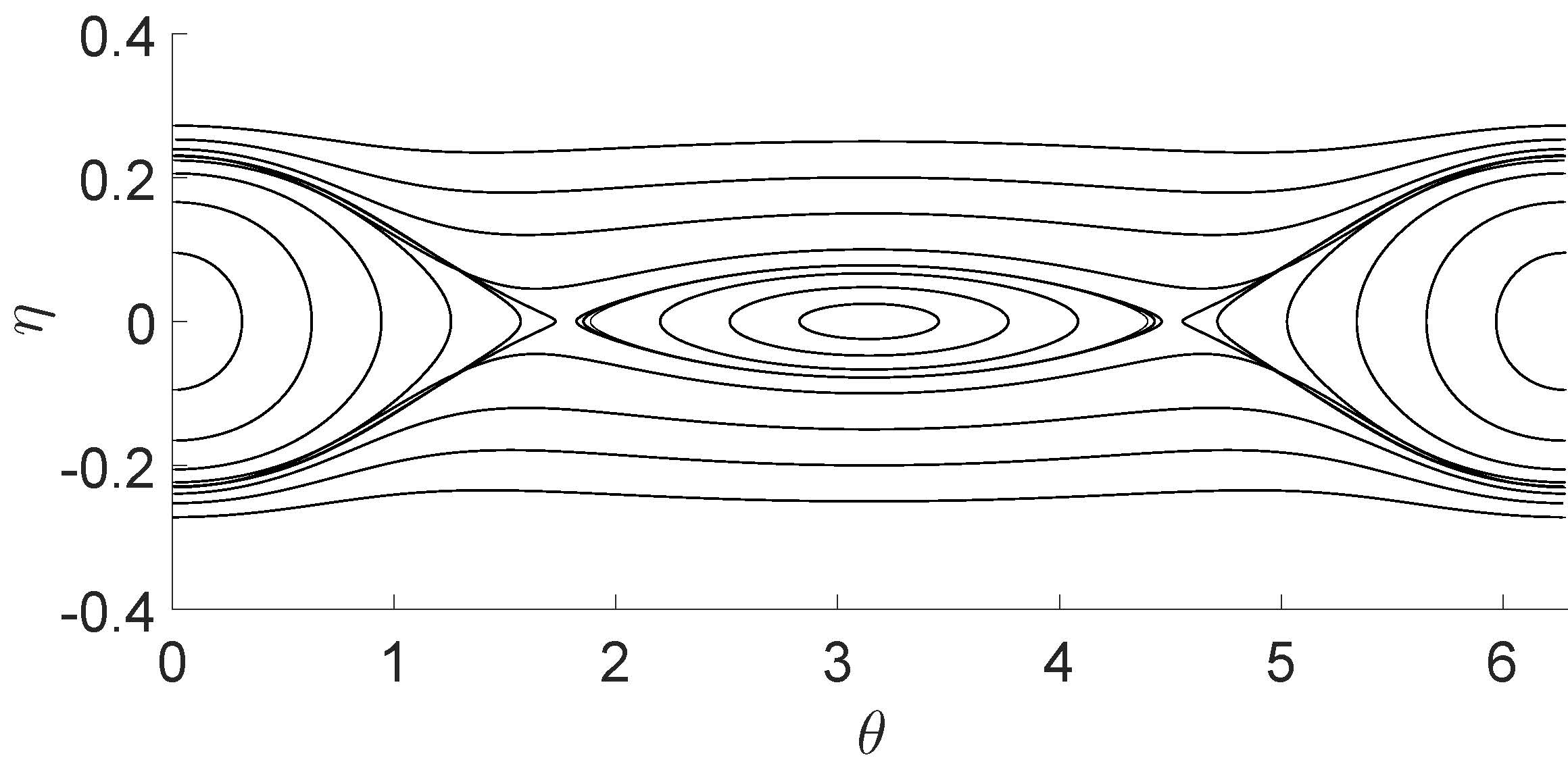}\\
(c)$I(0)=0.32$
\end{center}
\end{minipage}
\begin{minipage}{0.23\textwidth}
\begin{center}
		\includegraphics[width=0.95\textwidth]{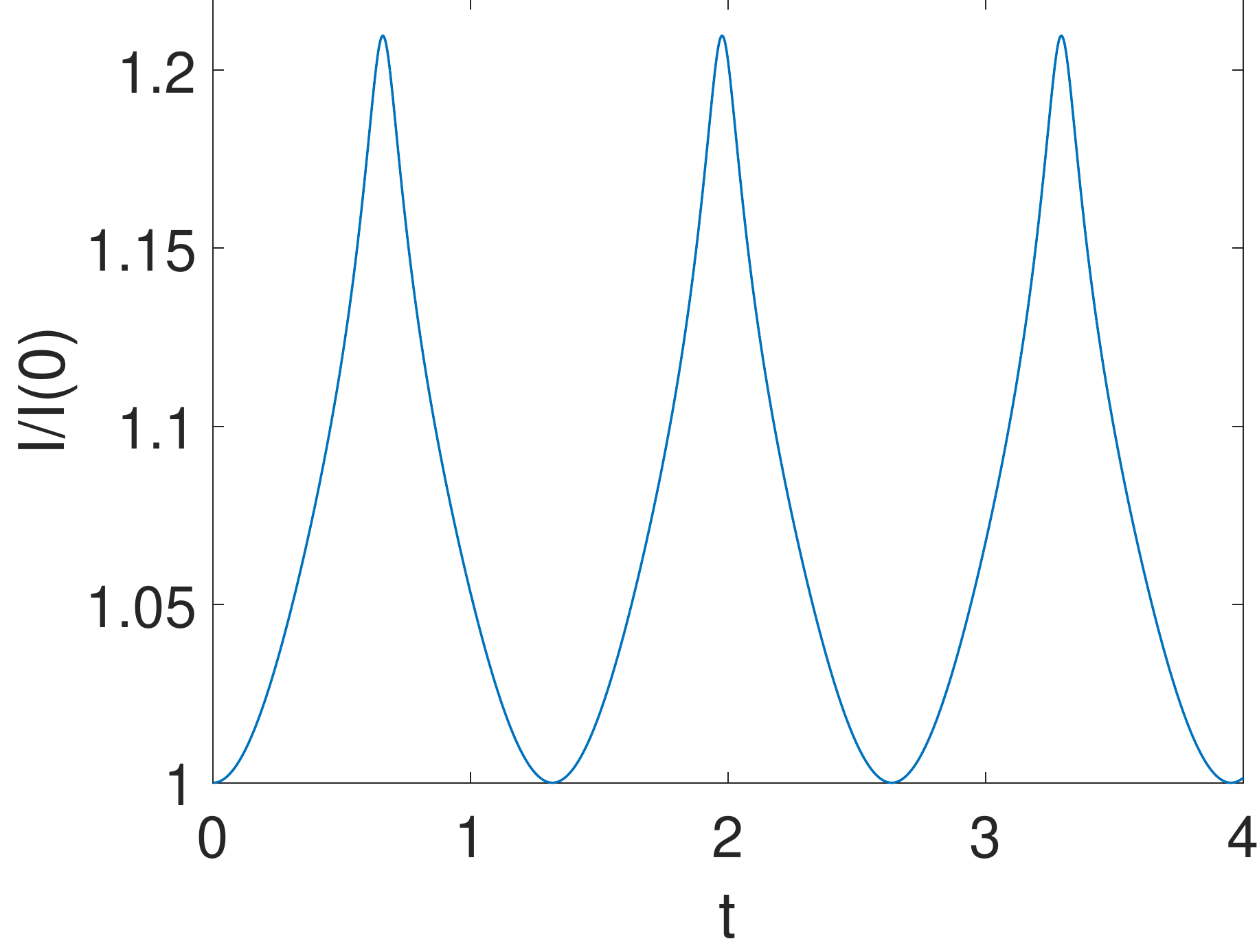}\\
(g)%$\frac{I(t)}{I(0)}$
\end{center}
\end{minipage}

\begin{minipage}{0.23\textwidth}
\begin{center}
	\includegraphics[width=0.95\textwidth]{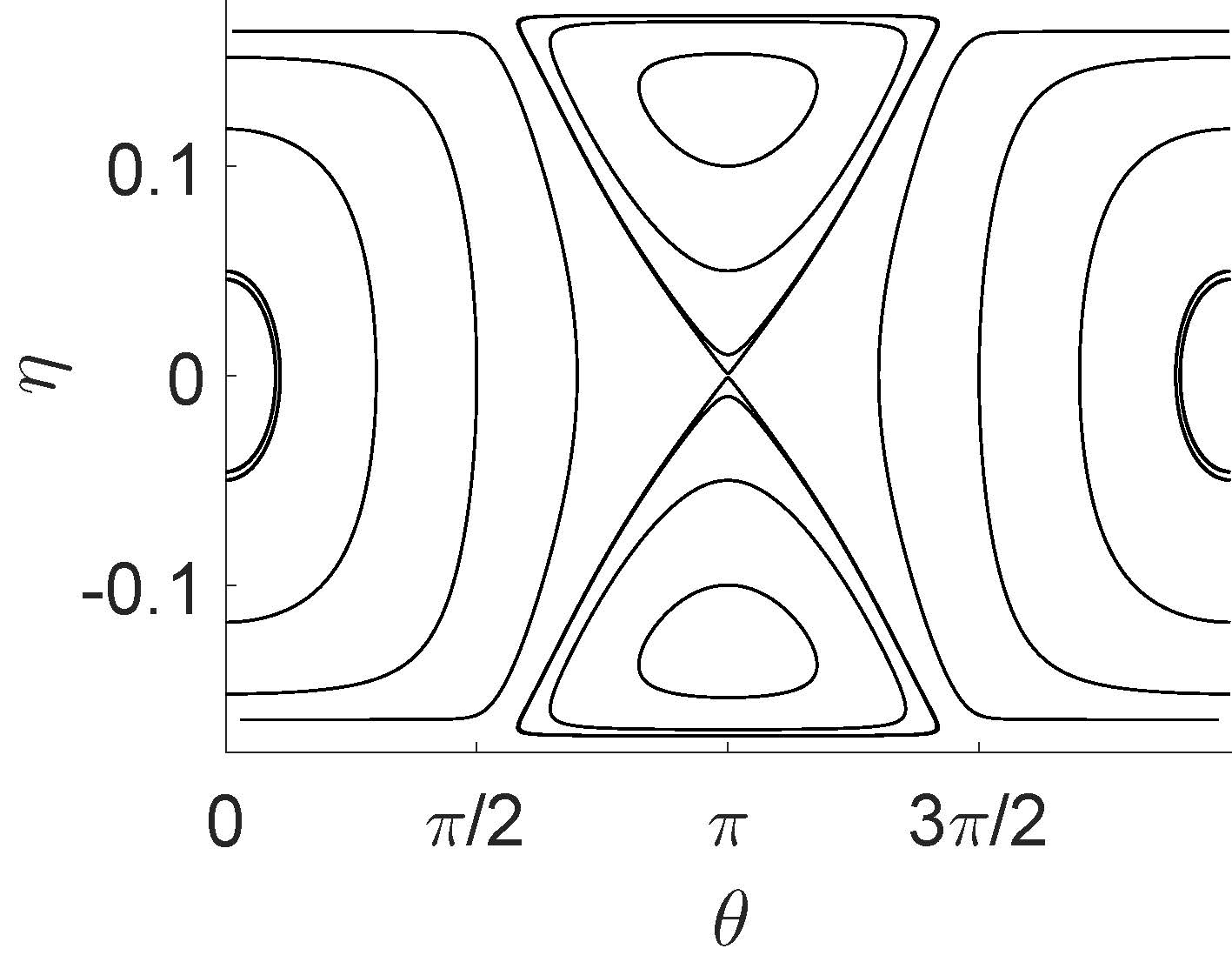}\\
(d)$I(0)=0.15$
\end{center}
\end{minipage}
\begin{minipage}{0.23\textwidth}
\begin{center}
		\includegraphics[width=0.95\textwidth]{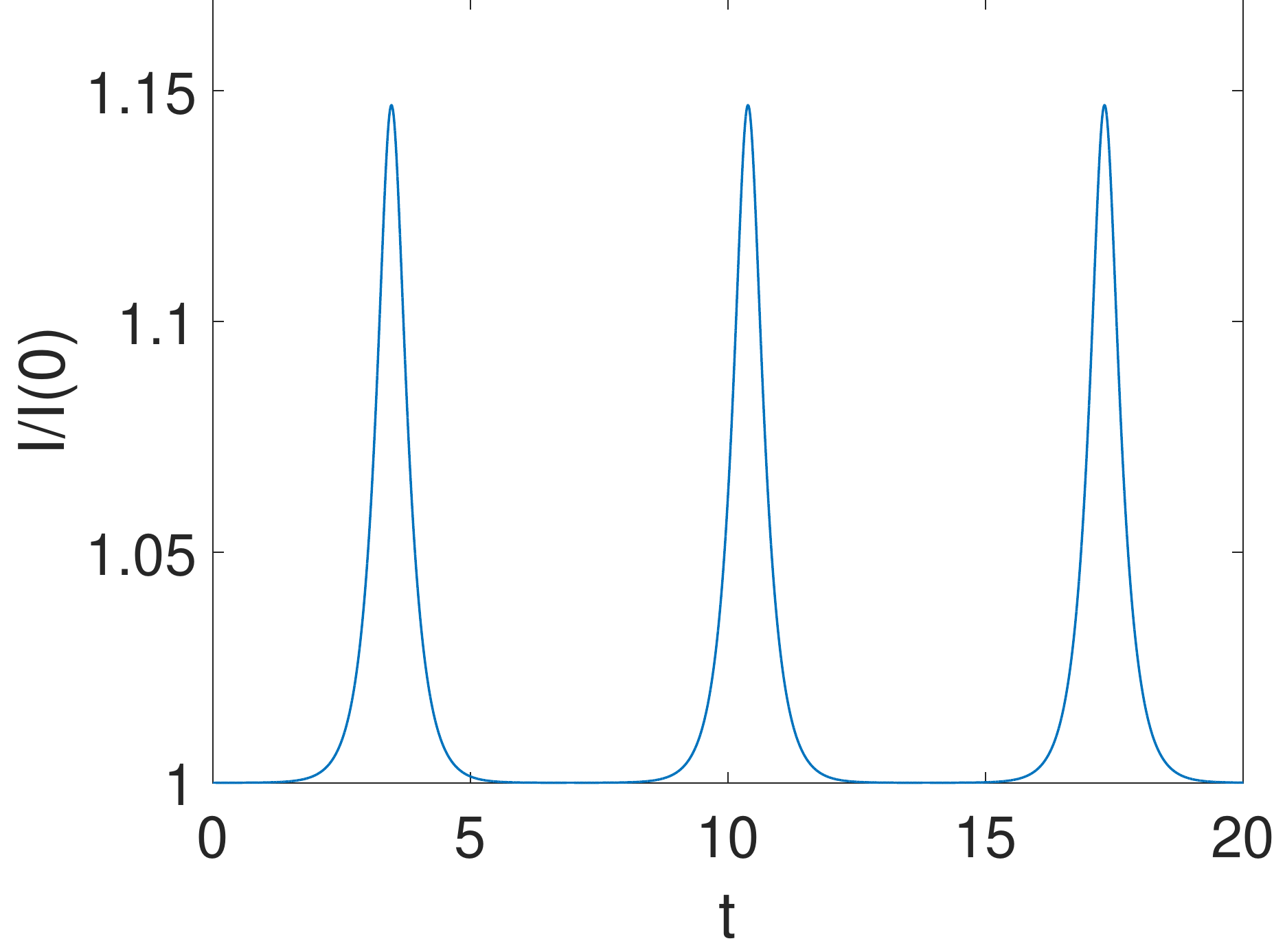}\\
(h)%$\frac{I(t)}{I(0)}$
\end{center}
\end{minipage}

\caption{First column : Pseudo phase portraits of $\eta$ vs $\theta$. Second column: Plots of variations of $I(t)$ normalized by $I(0)$; $\frac{I(t)}{I(0)}$. (a)-(e) $I(0) = 0.1$, (b)-(f) $I(0) = 0.7$, (c)-(g) $I(0) = 0.32$ and (d)-(f) $I(0) = 0.15$. }\label{fig:pp}
\end{figure}

Unlike \cite{Kimura1988} for the problem of two point vortices in a circular domain, $I$ is no longer conserved.  The solutions to \eqref{eq:rotlet_eqs2} obtained through a numerical simulation for a variety of initial values of $I(0)$, $\eta(0)$ and $\theta(0)$ can be plotted as a pseudo phase portrait of $\eta(t)$ vs $\theta(t)$. Three distinct pseudo phase portraits are seen as the parameter $I(0)$ varies; with the approximate intervals of $I$ found from numerics being $I<0.119$ fig. \ref{fig:pp}(a); for $0.119<I<0.22$ fig. \ref{fig:pp}(d) and for $I>0.22$ fig. \ref{fig:pp}(b)-(c).   These are not true phase portraits since  each apparent orbit  actually fills out a small area. This is true even for the nearly Hamiltonian cases of small $I(0)$ and large $I(0)$ shown in figs. \ref{fig:pp}(a)-(b), while in fig. \ref{fig:pp}(c), the trajectories show clear intersections. The variations in $I(t)$ show a similar pattern. When $I(0) = 0.32$ an intermediate value $I(t)$ shows large variations, fig. \ref{fig:pp}(c),  while in the other two nearly Hamiltonian cases, $I(t)$ shows small variations, fig. \ref{fig:pp}(a)-(b). The second column in fig. \ref{fig:pp} shows the variations in $I(t)$ for initial conditions of $(\eta(0), \theta(0))$ that produce the large variations in $I(t)$ for each corresponding case of $I(0)$. These large variations occur in the region of the homoclinic and heteroclinic trajectories of the apparent saddle type fixed points in fig. \ref{fig:pp}(a)-(c). For small $I(0)$, the pseudo phase portrait shown in fig. \ref{fig:pp}(a) is nearly identical to the phase portrait of the two vortex system, shown in \cite{Kimura1988}, while for the range $0.119<I<0.22$ the pseudo phase portrait fig.\ref{fig:pp}(d) is topologically similar to the one case of two point vortices but in the range where $0.21374<I<0.29038$. The pseudo portraits for $I>0.22$ shown in fig.\ref{fig:pp}(b)-(c) are unique to the case of the two rotlets. These pseudo phase portraits provide a basis for classifying the mixing properties of the fluid in the domain due to the motion of the rotlets.

\section{Mixing of fluid due to a pair of rotlets in a circular domain}
The two rotors act as stirrers whose location in the domain varies quasiperioically. To investigate mixing of the fluid in the circular domain, the equations of motion of the rotlets, \eqref{eq:rotlet_circle}, have to be augmented by the advection equation for a fluid tracer that is not collocated with any of the $N$ rotlets. A fluid tracer is advected with a velocity that is a superposition of the velocities induced by each rotlet and each of the corresponding image singularities. If the location of a fluid tracer is denoted by $(x,y)$, then the velocity $(u_x, u_y)$ of the fluid tracer is given by the advection equation 
\begin{eqnarray}\label{eq:tracer}
 u_x = \sum_{i=1, }^N\frac{\partial \psi_i}{\partial y} \;\;\;\;\; 
 u_y = \sum_{i=1, }^N-\frac{\partial \psi_i}{\partial x}.
\end{eqnarray}

Two  interpretations are possible for the equations of advection of a fluid particle in the domain. The advection equations \eqref{eq:tracer} can model a two dimensional time dependent dynamical system, where the explicit time dependence is through the positions of the $N$ rotlets, which can be obtained from \eqref{eq:rotlet_circle}. Following this the trajectories of a fluid particle are allowed to self-intersect in the domain. A second possible interpretation is that equations \eqref{eq:rotlet_circle}, \eqref{eq:tracer}  define a time independent $2N+2$ dimensional dynamical system, a restricted planar 3-rotlet system.  We will adopt the first view and treat the advection of fluid particles as a two dimensional time dependent dynamical system, where the time dependence comes from the quasiperiodic motion of the rotors. The velocity of the fluid in the circular domain is then defined by the vector field $(u_x,u_y)$. Let the time dependent flow map for this dynamical system be denoted by $\phi_{t_0}^t(x(t_0), y(t_0)) : \mapsto (x(t), y(t))$. The flow map $\phi_{t_0}^t$  takes as input an initial condition $(x_{t_0}, y_{t_0})$ and maps it to the solution of the dynamical system at time $t$.

It is easy to see that mixing of the fluid is poor in the special case when $N=1$. A single rotlet in the domain moves only due to the influence of the boundary, i.e., its advection is due to the image singularity system alone. The velocity of the rotlet $(\dot{x}_1, \dot{y}_1)$ given by \eqref{eq:rotlet_circle} is Hamiltonian and the quantity $I=x_1^2+y_1^2$ is conserved. The rotlet moves along a circle with a constant speed and its motion is periodic. The dynamics of a rotlet and a tracer particle in the fluid therefore reduce to a restricted two rotlet case. The velocity of a tracer given by \eqref{eq:tracer} is periodic (but nonlinear) and no mixing of the fluid occurs in a large part of domain.

%\subsection{Chaotic mixing by two same spin rotlets}
When two rotlets  are present in the circular domain, more complex dynamics of the rotlets as well as the fluid particles ensue. We restrict our attention to the simplest case of two same spin rotlets, which is the minimal case where fluid mixing is possible. The same spin is motivated by the practical realization of identical spinning magnetic microspheres whose spin velocities generated by a magnetic field are equal. These mixing dynamics depend upon the region of the fluid that the rotlets visit. Obviously the dynamics of the rotlets, \eqref{eq:rotlet_circle} themselves depend on the distance between the two rotlets, $d$ and distance between each rotlet and the center of the circle, $r_1$ and $r_2$. This large parameter space is numerically explored by using the underlying nearly Hamiltonian structure of the system and the three qualitatively distinct pseudo phase portraits. Initial conditions for the system are chosen in terms of $\eta$ and $\theta$ for different initial values $I(0)$.

%To investigate mixing of the fluid, we first define a blob of a fluid at initial time $t_0$, as an  open set, $B(t_0)$, in the circular domain not containing any rotlet. The initial blob is mapped to the set $B(t) = \phi_{t_0}^t(B(t_0))$, such that every point $(x(t_0), y(t_0) \in B(t_0)$ is mapped to $(x(t), y(t)) = \phi_{t_0}^t(x(t_0), y(t_0))$. An analytical solution of the flow map is nontrivial and we use numerical  simulations of \eqref{eq:tracer} to investigate the evolution of arbitrary blobs of fluid.

An analytical solution of the flow map is not possible and we use numerical  simulations of \eqref{eq:tracer} to investigate the evolution of arbitrary blobs of fluid. The mixing of the fluid  is quantified by a Shannon entropy field, which has also been referred to as locational entropy \cite{Camesasca_2006, pushpavanam_pre2017}. In spirit and in formal computation the locational entropy is the same as finite time entropy field \cite{froyland_fte_2012}. An entropy field is the locational entropy associated with different subsets of the domain which can identify regions of good (or fast) mixing and regions of poor (or slow) mixing. This is better suited to the current problem because the dynamics of tracer particles do not have a common rational time period in the entire domain, precluding the use Poincare maps. The entropy field is calculated as follows.

The domain is divided in $n$ subsets denoted by $\{B_1, B_2, ..., B_n\}$. From the perspective of studying mixing, each subset initially contains a  species of particles unique to that box, with a total of $n$ species of particles in the entire domain. The advection of the fraction of the fluid material from a box $B_i$ to box $B_j$ is 
\begin{equation}\label{eq:p_ij}
    p_{ij} = \frac{\mu(\phi_{t_0}^t(B_i)\cap B_j)}{\mu(B_i)}
\end{equation}
where $\mu$ denotes the Lebesgue measure.  Equation $p_{ij}$ can also be interpreted as the probability that the set $B_i$ will be mapped to into set $B_j$ by the flow map $\phi_{t_0}^t$. In computational approximation of this probability, each subset $B_i$ is chosen to be of an equal area and each subset is seeded with the same number of fluid particles.  The locational entropy is defined as
\begin{equation}
    S_j = -\sum_{i=1}^n p_{ij}\ln{p_{ij}}
\end{equation}
if $p_{ij} \neq 0$ and zero otherwise. The normalized total entropy is 
\begin{equation}
    \overline{S} = \frac{1}{\ln{n}}\sum_{j=1}^n S_j.
\end{equation}
The total entropy is normalized such that when $\overline{S}=1$ perfect mixing occurs, with any subset in the fluid domain containing equal fractions of all the species of particles.  Regions of the fluid domain with low  locational entropy identify sets that are poorly mixed.

\subsection{Numerical Simulations}
In our simulations, we chose the number of subsets (i.e. species of particles) to be $n=225$ and the number of particles of each species to be $240$. A wide range of initial conditions for the rotlets were chosen to investigate the mixing of the fluid due to their motion. The  initial values of the quantity $I(0)$ of the rotlet-pair were chosen in a large range to represent the dynamics for each type of pseduo phase portrait shown in fig. \ref{fig:pp}. For each value of $I(0)$, the initial conditions $(\theta(0), \eta(0))$ spanned all the qualitatively different level sets shown in fig.\ref{fig:pp}. The total entropy $S$ for each case was computed along with the locational entropy to identify the subsets of the four dimensional phase space (of the two rotlet system) parameterized by $(I(0), \theta(0), \eta(0))$ that produce good and poor mixing.

\begin{figure*}
\begin{center}
		\includegraphics[width=0.95\textwidth]{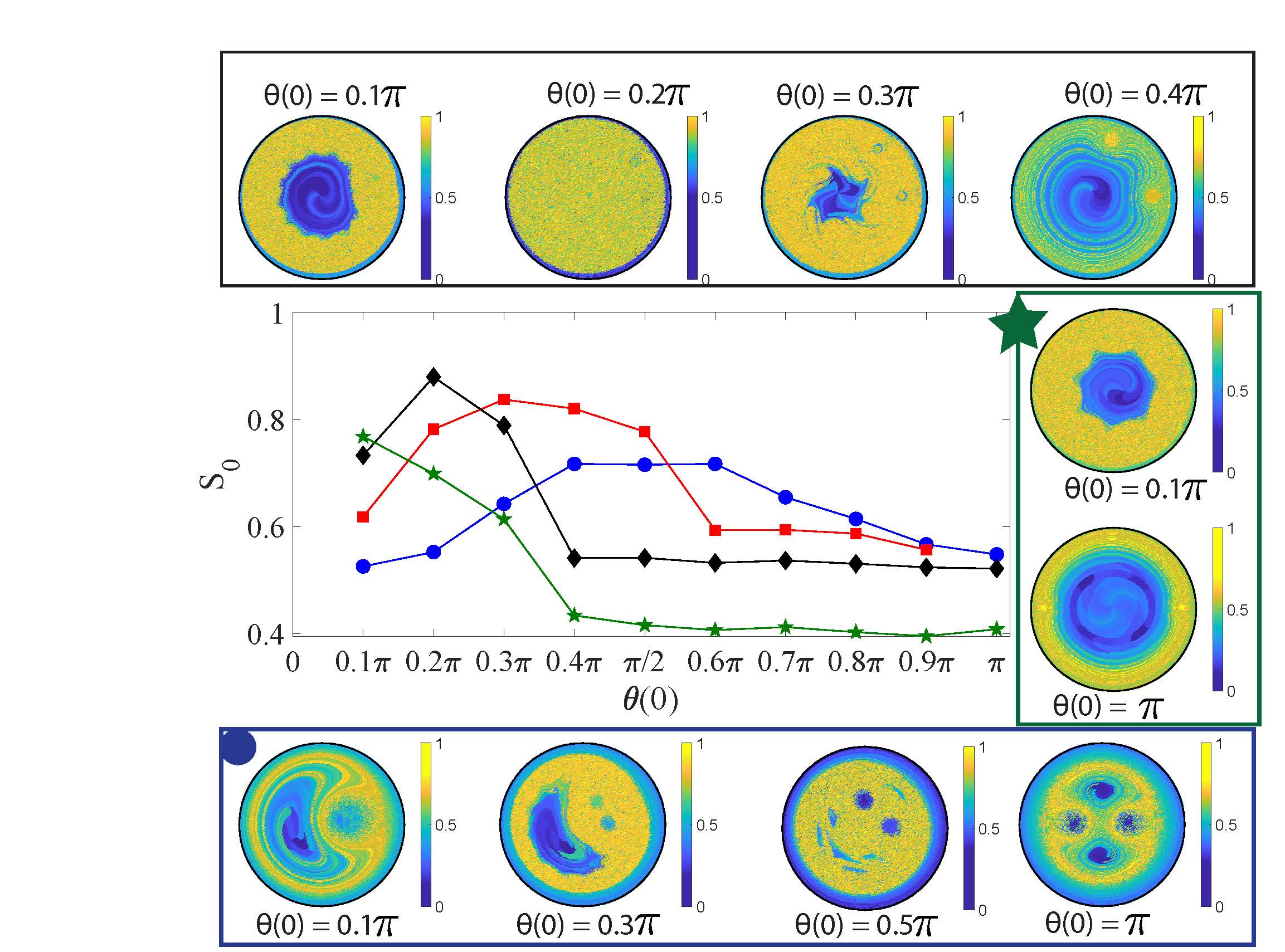}\\
\end{center}
\caption{Entropy $S_0$ and the entropy field. The graph in the center shows the entropy $\overline{S}$ for five different values of $I(0)$  with varying $\theta(0)$ and in each case $\eta(0) = 0$. The images around the graph show the entropy field for a few initial configurations of the rotlets, with the images in the top row for the case $I(0)=0.5$, in the bottom row for $I(0)=0.1$ and the right column for $I(0)=0.7$.}\label{fig:S_field1}
\end{figure*}

\begin{figure*}
\begin{center}
		\includegraphics[width=0.75\textwidth]{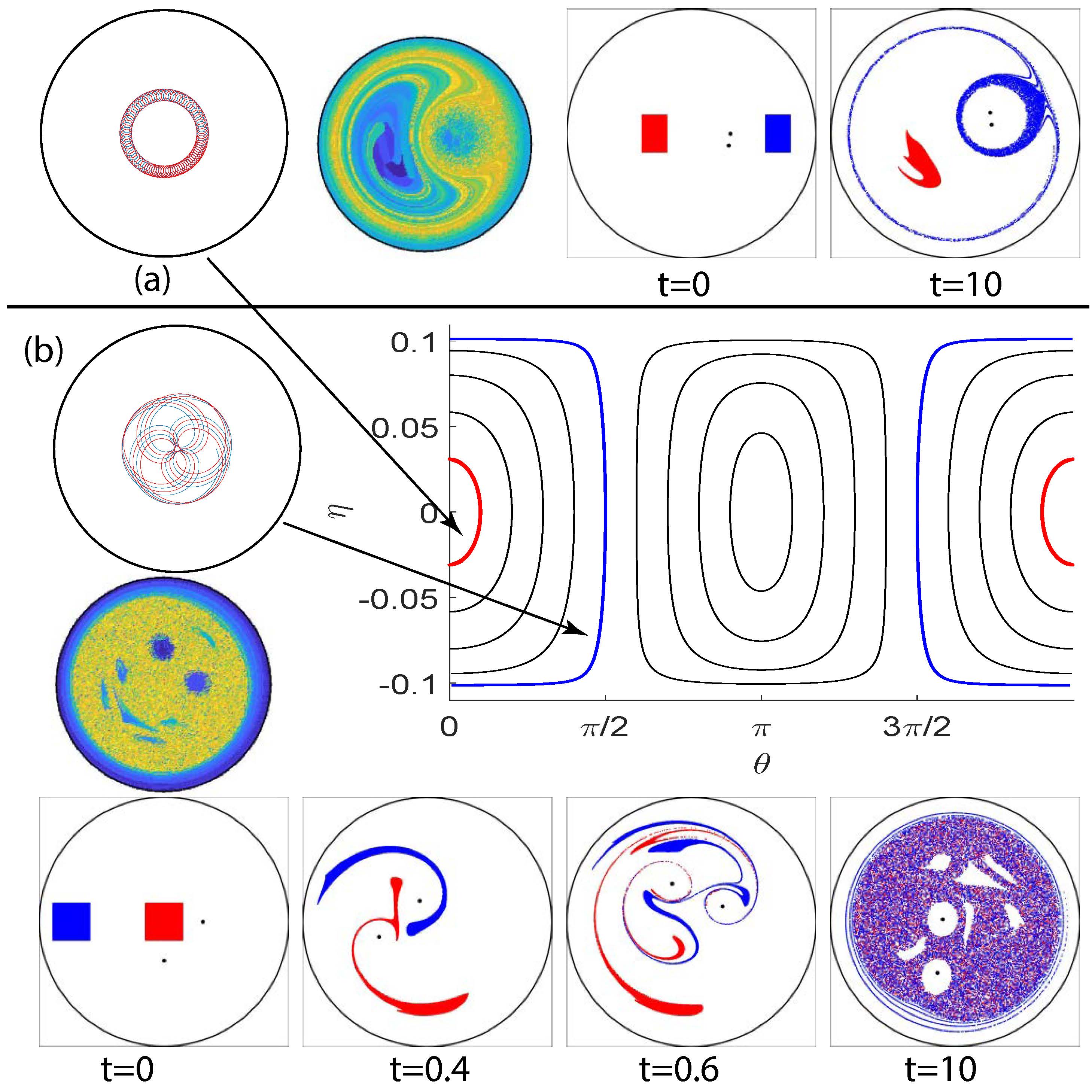}\\
\end{center}
\caption{Mixing dynamics for $I(0)=0.1$. (a) The initial conditions are $(\theta(0)=0.1\pi, \eta(0)=0)$. The top panel shows the trajectories of the two rotlets, the locational entropy plot at $t=10$ and the poor mixing of two blobs of fluid. (b) The initial conditions are $(\theta(0)=\pi/2, \eta(0)=0)$. The left and bottom panels show the trajectories of the rotlets, the locational entropy plot and the motion and mixing of two blobs of fluid in the time interval $[0,10]$. On the pseudo phase protrait in the center the orbits $(\theta(t), \eta(t))$ are shown for both the cases. The interior (red) orbit is for case (a) and the outer (blue) orbit is for case (b).}\label{fig:mixing_I01}
\end{figure*}

The graphs in the center of the fig. \ref{fig:S_field1} show the variation in the entropy $\overline{S}$ for different initial positions of the rotlets. These initial positions are categorized based on the initial values of $I(0)$, $\theta(0)$ and $\eta(0)$. Here it is once again emphasized that the parameter $I(0)$ is not conserved but along with the pseudo phase portraits are nevertheless useful to understand the mixing of the fluid. The parameter $\eta(0)$ quantifies the initial asymmetry in the distance of each rotlet from the center. Plots of the entropy field are shown for some initial conditions of $(I(0), \theta(0), \eta(0))$. The graph of $\overline{S}$ and the plots of the locational entropy $S$ show that in each case the total entropy $\overline{S}$ and the well mixed region increases in area initially with $\theta(0)$ and then eventually decreases.  For larger values of $I(0)$ the decrease in entropy begins to occur for smaller values of $\theta(0)$. This behavior can be better understood by examining the pseudo phase portraits of the interaction of the two rotlets. When $I(0) = 0.1$, the entropy rises until about $\theta \approx \pi/2$ for $\eta(0)$ and then decays. In the pseudo phase portrait as $\theta(0)$ increases from zero, the range of $\eta(t)$ also increases. Figure \ref{fig:mixing_I01} shows the dynamics of the rotlets and the mixing produced by them for two different cases of $\theta(0)$. In case (a) the value of $\eta(t)$ varies between $-0.0302$ and $0.0302$, i.e the distance of each of the rotlets from the center vary between $\sqrt{0.0698} \approx 0.2642$ and $\sqrt{0.1302}\approx 0.3608$. The velocity of each rotlet is dominated by the presence of the other rotlet with the influence of the image systems being negligible. The trajectories of two rotlets always stay within a small annulus  as shown in fig. \ref{fig:mixing_I01}(a) leading to a region of poor mixing shown in the locational entropy plot in the top panel of fig. \ref{fig:mixing_I01}. A blob of fluid initially located within this poorly mixed region does not experience significant stretching and folding and stays poorly mixed with the rest of the fluid. A blob of fluid starting in the high locational entropy region does stretch, fold and spread throughout a thin outer region but this is insufficient to achieve high values of total entropy $\overline{S}$. Conversely when the initial conditions are $(\theta = 0.4\pi, \eta=0)$, the orbit in the pseudo phase portrait shows a large variation in $\eta$ from almost $-0.1$ to $0.1$. This means that the distance of the two rotlets from center varies of zero to $\sqrt{0.2}\approx 0.4472$; with the orbits of the rotlets thus covering a larger area within the circular domain. This allows the fluid to be more uniformly mixed throughout a central core of the domain, except for some unmixed islands, producing a region with a high location entropy. The lower panel in fig. \ref{fig:mixing_I01} shows the mixing of two blobs of fluid, that experience rapid stretching and folding and mixing in this central region. Here it is worth emphasizing that the locational entropy plots show the entropy field computed at $t=10$ assigned to spatial locations fixed at initial time. The entropy field moves (in a lagrangian sense) with the rotlets and hence the blob plots in the lower panel show the unmixed island at locations different from that in the locational entropy plot.

\begin{figure*}
\begin{center}
		\includegraphics[width=0.75\textwidth]{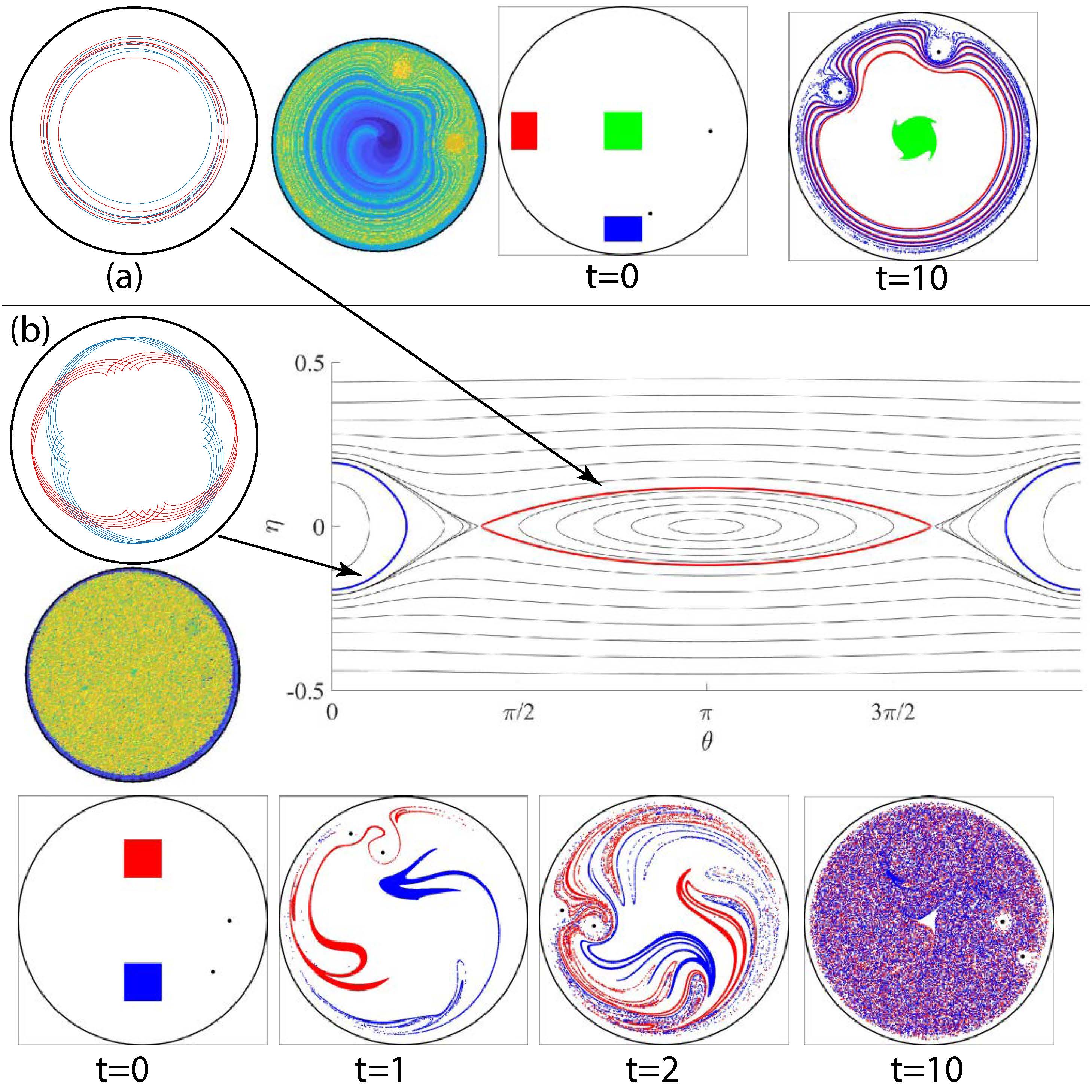}\\
\end{center}
\caption{Mixing for $I(0)=0.5$.  (a) The initial conditions are $(\theta(0)=0.2\pi, \eta(0)=0)$. The top panel shows the trajectories of the two rotlets, the locational entropy plot at $t=10$ and the poor mixing of the interior (green) blob of fluid with the blobs closer to the boundary (red and blue). The two outer blobs mix well in a thin outer annulus. (b) The initial conditions are $(\theta(0)=0.4\pi, \eta(0)=0)$. The left and bottom panels show the trajectories of the rotlets, the locational entropy plot and the motion and mixing of two blobs of fluid in the time interval $[0,10]$. On the pseudo phase protrait in the center the orbits $(\theta(t), \eta(t))$ are shown for both the cases. The interior (red) orbit is for case (a) and the outer (blue) orbit is for case (b).}\label{fig:mixing_I05}
\end{figure*}

For larger values of $I(0)$ the pseudo phase portrait is different and the entropy $\overline{S}$ shows an initial increase with $\theta(0)$ and then decreases for larger values of $\theta(0)$. Figure. \ref{fig:mixing_I05} shows the mixing in two cases of $\theta(0)$ for $I(0) = 0.5$. In case (a) $\theta(0) = 0.4\pi$ and the trajectory of the rotlets is confined to a thin annulus. In the pseduo phase portrait, the orbit for this case is shown in red, with $\eta(t)$ varying very slowly for large changes in $\theta(t)$. Essentially the rotlets never get close to each other. These dynamics produce slow mixing in an outer annulus with the locational entropy in this band being almost $0.5$ but leaves a large central region poorly mixed. The blob plots in the top panel show the motion of three blobs, with the blobs in the outer region mixing with each other, but the blob in the central region remaining coherent with very little deformation. In case (b) when $\theta(0)=0.2\pi$, $\eta(t)$ shows larger variation but over a smaller change in $\theta(t)$. The rotlets move close to and away from each other in a dance as the trajectories in fig. \ref{fig:mixing_I05}(b) show producing a well mixed region through out the domain except for a thin outer region. The lower panel in fig. \ref{fig:mixing_I05}(b) shows the dynamics of two blobs for this case, as the rotlets move in and away from each other material lines from these blobs are wrapped around both rotlets resulting in rapid stretching and folding.  As $I(0$ increases the graph of entropy (fig. \ref{fig:S_field1}) $\overline{S}$ shows a decline for smaller values of $\theta(0)$. This is because of the saddle like points lying on the $\eta=0$ line in the pseudo phase portraits move towards $\theta=0$ and $\theta=2\pi$ as $I(0)$ increases. Orbits not enclosing $\theta=0$ lead to rotlet motion and mixing dynamics qualitatively similar to case (a) in fig. \ref{fig:mixing_I05}, explaining the decline in $\overline{S}$ for larger values of $\theta(0)$.

\begin{figure}
\begin{center}
		\includegraphics[width=0.5\textwidth]{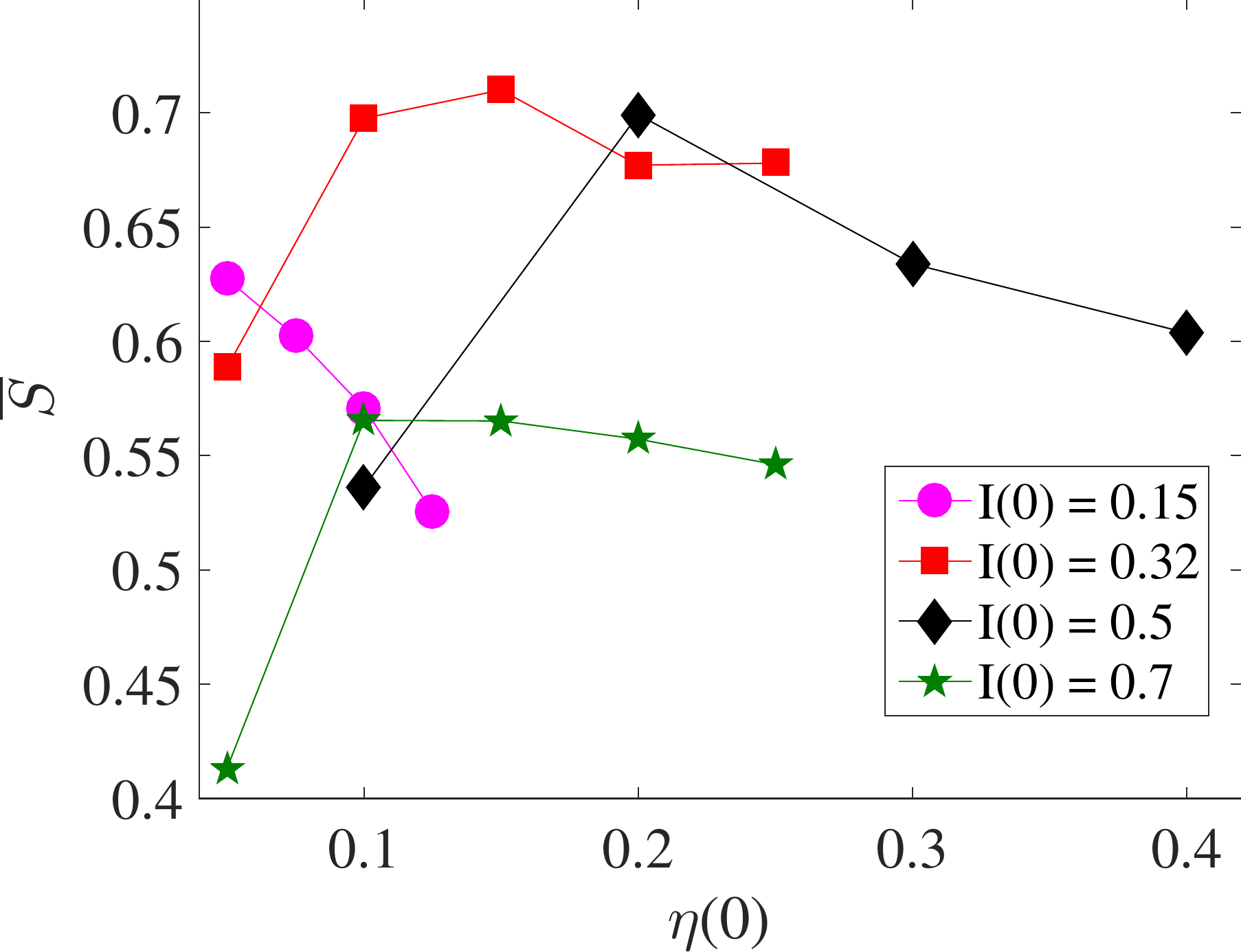}\\
\end{center}
\caption{Entropy $\overline{S}$ for $I(0)>0.119$. Initial conditions for all the cases are $(\theta(0) = \pi, \eta(0)>0)$. The initial value $\eta(0)$ is shown on the horizontal axis.}\label{fig:S_eta}
\end{figure}

One qualitative set of initial conditions of rotlets remain whose mixing dynamics  is not explored in fig. \ref{fig:S_field1}. For $I>0.119$ initial conditions $(\theta = \pi,\eta \neq 0)$ produce qualitatively different trajectories of the rotlets. When $0.119<I(0)<0.22$ and $(\theta(0)=\pi, 0<|\eta(0)|\leq I(0))$, then $\eta(t)$ never changes sign. As the pseudo phase portrait for this case, fig. \ref{fig:pp}(d), shows as $|\eta(0)|$ increases the variation in the distance of the rotlets from the center decreases, decreasing the area covered by the rotlets and the entropy $\overline{S}$. This trend can be observed in fig. \ref{fig:S_eta}.
When $I(0)>0.22$, the pseudo phase portraits have center type fixed point at $(\theta = \pi, \eta=0)$, around which nearly closed orbits exist within a region bounded by heteroclinic trajectories connecting the saddle points, see fig. \ref{fig:pp}(b)-(c). Longer orbits around the center type fixed point produce better mixing. The entropy therefore increases initially with increasing $\eta(0)$. However once the initial conditions cross the heteroclinic trajectories, the motion of the rotlets is such that $\eta(t)$ does not change sign; in fact in such cases $\eta(t)$ changes relatively very little for the entire range of $0\leq \theta(t)\leq 2\pi$. The rotlet trajectories in the circle are such that they remain at a nearly constant distance from the center.  At closest approach of the two rotlets, when $\theta = 0$ (or $\theta = 2\pi$) is $r_2^2-r_1^2 = I+\eta - (I-\eta) = 2\eta$. As $\eta(0)$ increases, $min(r_2^2-r_1^2)$ approaches $2\eta(0)$, i.e. even at closest approach the distance between the rotlets is still large and little fluid is pushed from the innner rotlet to the outer rotlet (or vice versa). Figure \ref{fig:S_eta} shows this decreasing entropy for larger values of $\eta(0)$.

\section{Conclusion}
The interactive dynamics of  two micro rotors modeled as two rotlets in a Stokes fluid in confined circular domain can act as mobile micromixers.  The dynamics of the rotlets themselves are no longer Hamiltonian and the angular impulse of the system is not conserved. For almost all initial conditions of the rotlets, the motion of the fluid is such that there are regions of the domain with chaotic mixing. For subsets of initial conditions of rotlets, the region of chaotic mixing of the fluid is very large approaching the entire domain. The paper presents a complete parametric investigation of the mixing of the fluid and identified the qualitative trends in the locational entropy field and the overall entropy of the fluid due to the chaotic advection of the fluid. These qualitative trends are related to the pseudo phase portraits of the system, that is made possible by the fact the two-rotlet system is nearly Hamiltonian.

The mixing dynamics due to the two rotlet system can have an advantageous physical realization over the case of the well known blinking vortices (or blinking rotlets). Blinking singularities require physical stirrers with piecewise continuous (in time) operation. The mobile rotlets investigated in this paper have smooth motion and can in theory be placed anywhere in the domain. In recent work \cite{bfst_dscc2018, bft_ijira2018}, the authors showed the possibility of steering two spinning spheres with a single control input (such as a magnetic field) to any desired location in a circular domain. Such techniques can be useful to create microrotors as controllable mobile micromixers, whose initial positions can be steered to realize orbits in any of the pseudo phase portraits.  The work presented in this paper illuminates where such microrotors should be steered to initially to produce fast mixing of the fluid. 

%\bibliographystyle{unsrt}
%\bibliography{Mixing}

%%%%%%%%%%%%%

%%%%%%%%%%%%%%%%%%%
\end{document}